\documentclass[useAMS,usenatbib,a4paper,referee]{mn2e}
\usepackage{graphicx,times}
\usepackage{amsmath}
\usepackage{longtable}

\title[Multi-wavelength observations of polarized blazars]
{Constraining Einstein's Equivalence Principle With Multi-Wavelength Observations of Polarized Blazars}

\author[Yi et al.]
{Shuang-Xi Yi$^{1}$\thanks{yisx2015@qfnu.edu.cn}, Yuan-Chuan Zou$^{2}$\thanks{zouyc@hust.edu.cn}, Xuan Yang$^{2, 3}$, Bin Liao$^{2}$, Shao-Wen Wei$^{4}$\\
 $^1$School of Physics and Physical Engineering, Qufu Normal University, Qufu 273165, China\\
 $^2$School of Physics, Huazhong University of Science and Technology, Wuhan 430074, China\\
 $^3$Purple Mountain Observatory, Chinese Academy of Sciences, Nanjing, 210008, China \\
 $^4$Institute of Theoretical Physics \& Research Center of Gravitation, Lanzhou University, Lanzhou 730000, China.\\}

\begin{document}

%\date{Accepted 1988 December 15. Received 1988 December 14; in original form 1988 October 11}

%\pagerange{\pageref{firstpage}--\pageref{lastpage}} \pubyear{2012}
%
%\label{firstpage}

\maketitle
\begin{abstract}
In this paper, we present a novel method to test the Einstein's Equivalence Principle (EEP) using (simultaneous) multi-wavelength radio observations of polarized blazars. We analyze simultaneous multi-wavelength polarization observations of 3C 279 at 22, 43, and 86 GHz obtained by two antennas of the Korean VLBI Network. We obtained 15 groups of polarization data, and applied the Metropolis-Hastings Markov Chain (MHMC) to simulate the parameters when considering the EEP effect and the simplest form of Faraday rotation (single external Faraday screen). The final results show the constraint of the parameterized post-Newtonian (PPN) parameter $\gamma$ discrepancy as $\Delta \gamma_{p} = (1.91\pm0.34)\times10^{-20}$.  However, the single external Faraday screen is an oversimplification for blazars because there are
numerous observations show complex Faraday rotation behavior for blazars due to internal/external Faraday dispersion, beam depolarization, etc. The value $\Delta \gamma_{p}$ results of this paper can only be considered as upper limits. Only if all other effects are revealed and considered, should the result be taken as a direct measurement of the violation of the EEP.
\end{abstract}

\section{Introduction}\label{sec:intro}

The Einstein Equivalence Principle (EEP) is one of the cornerstones of general relativity and many other metric theories
of gravity. It states that the trajectory of any freely falling, uncharged test body is independent of its energy, composition,
or internal structure. In the theory of the parameterized post-Newtonian (PPN) approximation, the EEP requires that the PPN parameter $\gamma$ of two different
particles should be in agreement with
\begin{equation} \label{0}
  \Delta \gamma \equiv \gamma_a-\gamma_b=0,
\end{equation}
where $\gamma$ denotes the level of space curvature
per unit rest mass, and the subscripts a and b denote two different test particles (Misner et al. 1973; Will 2014). Therefore, two particles
should have the same time delay during the propagation from the emitting source to the observer.

However, the observed time delays of some observational cosmic messengers (e.g. photons, neutrinos, or
gravitational waves), or the same type of particles with different frequencies or the linearly polarized beams have been used to
test the EEP (Shapiro 1964; Krauss \& Tremaine 1988; Longo 1988; Toma et al. 2012; Yang et al. 2017; Wu et al. 2017; Wei \& Wu 2019).
Different works have taken different sources, e.g., delays between arrival times of photons and neutrinos from SN1987A (Krauss \& Tremaine 1988; Longo 1988), the photons in different energy bands of gamma-ray bursts (GRBs) (Gao et al. 2015; Yu et al. 2018; Sang et al. 2016), radio signals for different frequency bands of fast radio bursts (FRBs) (Wei et al. 2015; Nusser 2016), and the time lag between keV photons and TeV photons from TeV blazars (Wei et al. 2016). In other works, the Crab pulsar (Yang \& Zhang 2016; Zhang \& Gong 2017), a PeV-energy neutrino
from the blazar PKS B1424-418 (Wang et al. 2016), and gravitational wave (GW) sources (Wu et al. 2016; Kahya 2016) have also been used to constrain the EEP.

The most stringent constraint on the EEP was established by Yang et al. (2017), who tested the EEP with the parameter $\gamma$ by taking GRB 110721A with high linear polarization of the gamma-ray-band into consideration. However, because of the limitation of observations, the accuracy of the polarized data is not good enough.  The measurement of the polarization of GRB 110721A gave a constraint on the EEP with $\Delta \gamma_{p} < 1.6\times10^{-27}$. However, the polarization in the radio band is much easier to obtain. Here, we
present the constraint on the value of $\Delta \gamma_{p}$ through a new method with multi-wavelength polarization
observations of blazars at radio bands.

\section{Method of testing the EEP} \label{sec:method}
Adopting the PPN approximation, Shapiro time delay in a gravitational
potential $U({\bf r})$ is given by (Shapiro 1964; Krauss \& Tremaine 1988; Longo 1988)
\begin{equation}\label{1}
\delta t_{\rm gra} = -\frac{1+\gamma}{c^{3}}\int_{\bf r_{\rm e}}^{\bf r_{\rm o}}U(\bf r) d \bf r,
\end{equation}
where $\bf r_{\rm o}$ and $\bf r_{\rm e}$ denote the locations of the
observer and emitting source, respectively. We consider linearly polarized light, which is composed of two circularly polarized beams (labeled `r' and `l'). In general relativity, $\gamma=1$. If the EEP is violated, the $\gamma$ values of photons with right- and left-handed circular polarization should differ slightly, leading to the two circularly polarized beams passing through the same gravitational potential with different Shapiro time delays (Yang et al. 2017). The relative Shapiro
time delay is given by
\begin{equation}\label{2}
\Delta t_{\rm gra} = \mid\frac{\Delta \gamma_{\rm p}}{c^{3}}\int_{\bf r_{\rm e}}^{\bf r_{\rm o}}U(\bf r) d \bf r\mid,
\end{equation}
where $\Delta \gamma_{\rm p} \equiv \gamma_{\rm l}-\gamma_{\rm r}$ is the
difference of $\gamma$ for the left and right polarized light.
Any additional contributions from the intrinsic time lag, Lorentz invariance violation time delay and the special relativity time delay are considered negligible in this paper. $U({\bf r})$ is mainly the gravitational potential of the Milky Way galaxy, and
\begin{equation}\label{3}
 \int_{\bf r_{\rm e}}^{\bf r_{\rm o}}U(\bf r) d \bf r \simeq GM_{\rm MW}\ln \frac{d}{b},
\end{equation}
where $G=6.68 \times10^{-8}$$\rm erg$ $\rm cm$ $\rm g^{-2}$ is the gravitational constant, $M_{\rm MW}\approx 6\times10^{11}M_{\rm \odot}$ is the total mass of the Milky Way, $d$ is the distance from the Earth
to the source, and $b$ is the impact parameter of the rays.
The impact parameter can be calculated as
$b = r_{\rm G}\sqrt {1-\left(\sin\delta_{\rm S}\sin\delta_{\rm G}+\cos\delta_{\rm S}\cos\delta_{\rm G}\cos(\beta_{\rm S}-\beta_{\rm G})\right)^{2}}$, where $r_{\rm G}=8.3$ $\rm kpc$ denotes the radial distance from the Sun to the Galactic
center, $\beta_{\rm S}$ and $\delta_{\rm S}$ are the right ascension and
declination of the source in  equatorial coordinates, and ($\beta_{\rm G}=17^{h}45^{m}40.04^{s},\delta _{\rm G}=-29^{\circ}00^{\prime}28.1^{\prime\prime}$) are the coordinates of the Galactic center (Gao et al. 2015; Yang et al. 2017).
The proper distance $d=zc/H_0$, where $z$ is the redshift and $H_0=67.74$ km.s$^{-1}$.Mpc$^{-1}$ (Planck Collaboration et al. 2018).

Therefore, the rotation of the linear polarization angle during the propagation
from the emitting source to the observer is expressed as:
\begin{equation}\label{4}
\Delta \phi = \Delta t_{\rm gra}  \frac{ \pi c}{\lambda},
\end{equation}
where $\lambda$ is the wavelength.
However, the linear polarization angle can also be
modified by Faraday rotation when the light travels through magnetized plasma,
and the dependence of $\Delta \phi$ on the Faraday rotation is
$\Delta \phi\propto\lambda^2$ (Burn 1966), which is the simplest case of Faraday rotation but possibly an
oversimplification for blazars. Therefore, when considering both the Shapiro time delay
and Faraday rotation effects, the observed linear polarization angle
at wavelength $\lambda$  emitted from an astronomical transient event could
be expressed as
\begin{equation}\label{eq:6}
\phi_{\rm obs}( \lambda) = \phi_{\rm 0}+\frac{A}{\lambda}+B\lambda^2,
\end{equation}
where $\phi_{\rm 0}$ is the initial angle of the linearly polarized light, $A \equiv { \pi c} \Delta t_{\rm gra} $ represents the contribution from the Shapiro time delay effect, and $B$ is the rotation measure caused by Faraday rotation (Burn 1966).
The parameters $\phi_{0}$, A, and B
are constants to be determined from the fit to the observational data if there are linear
polarization measurements in several bands from the astronomical object, and the
time delay $\Delta t_{\rm gra} = \frac{A}{ \pi c}$ can be eventually obtained. Using the time lag $\Delta t_{\rm gra}$,
we then constrain the parameter $\Delta {\gamma}_{\rm p}$ between two beams with different
circular polarization.

\section{Tests of the EEP using Polarized Blazars} \label{sec:blazar}
We extensively searched for astrophysical objects with linear
polarization measurements in several bands, and we found a candidate with long-term multi-wavelength polarization observations, namely the powerful blazar 3C 279 with radio coordinates $\beta_{\rm S}= 12^{h} 56^{m} 11.1665^{s}, \delta _{\rm S}= -05^{\circ} 47^{\prime} 21.523^{\prime\prime}$ (Johnston et al. 1996; Kang et al. 2015). This blazar has a redshift of z=0.5362 (Marziani et al. 1995).  Kang et al. (2015) performed radio polarization observations using two 21-m telescopes of the Korean VLBI Network. Observations were carried out weekly in 35 sessions between 25 December 2013 and 11 January 2015, at 22, 43 , and 86 GHz. However, according to Kang
et al. (2015), there are 15 groups of the
polarized data in the three bands (22, 43, and 86 GHz) simultaneously (see Table 1)\footnote{ As mentioned in Table 2 of Kang et al. (2015), since they account only for the thermal noise of the receiver, the polarization angle errors are too small. The polarization angle values
should have a systematic uncertainty of approximately $2^\circ$ because they are calibrated using the Crab Nebula, which has a known polarization angle with this level of uncertainty. Therefore, an additional polarization angle error of $2^\circ$ is added in 15 groups of data in Table 1. }. Figure 1 shows the 15-point light curves of 3C 279 for the degree of linear polarization (top), linear polarization angle (middle), and flux density (bottom), at 22 GHz (black), 43 GHz (red), and
86 GHz (blue) from December 2013 to March 2015.

As shown in Eq. (6), the observed polarization angle $\phi_{\rm obs} (\lambda)$ is a function of the wavelength $\lambda$. The 15 groups of the polarized data for blazar 3C 279 are compiled with the observed polarization angles and their corresponding wavelengths.
For the nonlinear regression about the three polarized data of each group, we applied the function $ \phi_{\rm obs}(\lambda)$  of Eq. (6) to fit the observed data.
We used the Metropolis-Hastings Markov Chain (MHMC) to simulate the parameters. There are 31 parameters in total. Each set of data is simulated with
three parameters $A$, $B_i$, and $\phi_{\rm 0,\,i}$, where i varies from 1 to 15. All of the data use the same $A$, as the light path is the same, and consequently, the Shapiro time delay should be the same. We calculated the probability density for the fit results in parameter space for each group and multiplied them all to obtain the final probability density for all the 31 parameters. We show the contour plots of  $A$, $B_1$, $\phi_{01}$, $B_2$, and $\phi_{02}$ in Figure 2, and the other parameters are added in the Appendix with the contour plots. Values of all 31 fit parameters are shown in Table \ref{tab:2}. Figure 3 shows a group of the samples (the third group) with the observed linear polarization angles at different wavelengths. The red line is the best-fit line with Eq. (6), and the different components are shown in the figure with different shapes. This figure is an example of showing how different components contribute to the total polarization angle rotation.

The Shapiro time delay is $\Delta t_{\rm gra} = \frac{A}{ \pi c}=(7.21\pm1.29)\times10^{-13}$ s with the fit result $A=(0.068\pm0.012)$ cm. Taking all of the values into Eq. (3), we obtain the constraint of the $\gamma$ discrepancy as
\begin{equation} \label{6}
  \Delta \gamma_{p} = (1.91\pm0.34)\times10^{-20}.
\end{equation}

To avoid the source selection effect, we also searched for the other radio sources with the linear polarization measured in several bands, and found two bright sources with the linear polarization measured in 20, 8.6, and 4.8 GHz separately. The polarization angles are 10$^\circ$, -22$^\circ$, and -37$^\circ$ for PKS 0008-307 with redshift z=1.19 and  -36$^\circ$, -38$^\circ$, and -1$^\circ$ for  PKS 0008-264 with redshift z=1.096 at 20, 8.6 and 4.8 GHz, respectively. The right ascension and
declination for the two sources are ($\beta_{\rm S}= 00^{h} 10^{m} 35.92^{s}, \delta _{\rm S}= -30^{\circ} 27^{\prime} 48.3^{\prime\prime}$ ) and ($\beta_{\rm S}= 00^{h} 11^{m} 1.27^{s}, \delta _{\rm S}= -26^{\circ} 12^{\prime} 33.1^{\prime\prime}$ ). More details of the data can be seen in Massardi et al. (2008)\footnote{ Massardi et al. (2008) does not include an error for the polarization angle, but does include an error for the polarized flux P and flux density S. We applied the flux error to
estimate the error in the polarization angle. We can calculate $m$ from $m = P/S$ and its error $\Delta m$ from propagating $\Delta S$ and $\Delta P$. Then, the error $\Delta m$ is related to the error on the polarization angle $\Delta \phi$, like $\Delta \phi \approx \Delta m/2m$.}. The same fitting method is applied, and the fit results are shown in Figure 4 and Table 2. We obtain the constraint of the $\gamma$ discrepancy as $\Delta \gamma_{p} = (3.73\pm0.17)\times10^{-19}$ for PKS 0008-307, and the constraint of the $\gamma$ discrepancy as $\Delta \gamma_{p} = (2.40\pm1.40)\times10^{-19}$ for PKS 0008-264. The three values of $\Delta \gamma_p$ are different with order-of-magnitude
discrepancy. However, there are many types of Faraday rotation that can affect the results in this paper, such as, differential Faraday rotation, internal/external Faraday dispersion, or beam depolarization. These types may have different wavelength dependence, $\propto$ $\lambda$ or $\lambda^2$, or even whether it exhibits linear behavior (O'Sullivan et al. 2012; Karamanavis et al. 2016; Pasetto et al. 2016). Here we only consider the simplest scenario, in which the observed polarization angle $\phi_{\rm obs}$ is affected
by the effect of the simplest form of Faraday rotation ($\propto \lambda^2$) and Shapiro time delay ($\propto \lambda^{-1}$), and try to constrain the EEP using multi-wavelength radio polarization observations of blazars.
Besides, they were not simultaneously observed for different bands for PKS 0008-307 and PKS 0008-264.
This fact may introduce an additional uncertainty to the results.
Such variability may cause a variation of the Faraday rotation that
cannot be accounted for by the present model. Consequently, an extra
assumption is introduced for the analysis of these two additional sources. Therefore, the values of $\Delta \gamma_p$ should be taken as upper limits.

\section{Summary and discussion} \label{sec:con}

In this paper, we report a measurement of the time delay between lights with multi-wavelength polarization observations of blazars. The Faraday component and Shapiro time delay can be recovered with
the fit of the 15 groups of the multi-wavelength linear polarization angles at 22, 43, and 86 GHz with different time periods when using Eq. (6). We show the constraining of the $\gamma$ discrepancy as $\Delta \gamma_{p} = (1.91\pm0.34)\times10^{-20}$. In the ideal case, this could be taken as a direct measurement of the violation of the EEP using the linear polarization angles measured in several radio bands for one astrophysical object. However, it should be taken as an upper limit, since several underlying degenerate effects are not considered here. The result is model dependent in this paper, which is related to Shapiro time delay ($\propto \lambda^{-1}$) and the simple form of Faraday rotation
($\propto \lambda^2$). There are several models of the polarization, which can be realized by assuming that the observed radio emission is composed of two or more sources with different initial polarization angles and different RMs (O'Sullivan et al. 2012, Pasetto et al. 2016).
For example, as shown in Eq. (4) of Goldstein \& Reed (1984), they used two components to model the irregular Faraday rotation. Nevertheless, such expectations also rely on further assumptions like multiple compositions of the origin. Considering that there are other potential possibilities that cause the variation of the polarization angles, and for clarity, we do not carry out a detailed analysis with such models.
Therefore, our work could be regarded as a conceptual example demonstration of directly constraining the EEP from the frequency-dependent polarization angles.

We find it important to discuss the following point. Note that the upper limit from the polarization of GRB 110721A is $\Delta \gamma_{p} < 1.6\times10^{-27}$ (Yang et al. 2017), which is even smaller than the measured value reported in this Letter. Furthermore, these two results share a  similar method. There are two aspects of reasons, either on the side of $\gamma-$rays of the GRBs, or on the radio emission of the blazars. For the GRBs, the polarization might not be originating from the source, but rather obtained when the photons reach Earth, e.g., they pass through a magnetic field or suffer Compton scattering. For the blazars, it could be other effects that are degenerated with the EEP effect and dominates the angular rotation as shown in Eq. (6). The other effect could be the opacity effect, which was taken as a possible reason for the time delay of blazar flares in different radio bands (Karamanavis et al. 2016). This effect can also contribute to the polarization angle anomaly. If the emitting region is optically thick for the radio band, the opacity $\tau=1$ for different bands located at different places. For higher frequency, it locates deeper into the source. The emission suffers more magnetized plasma (i.e., larger RM), and consequently, the Faraday rotation is larger than the case of the same emitting region. As this effect also accompanies time delay, a careful modeling might be able to determine how much of the angular rotation results from the opacity effect. Another effect could be that radio emissions of different bands are mainly from different directions. Then, the polarization angles could be any values at different bands. This, in general, could be resolved with a more powerful telescope array with high angular resolution.
Consequently, the obtained value of $\Delta \gamma_{p}$ does not directly represent the violation of the EEP.
Therefore, even if we obtained a value of $\Delta \gamma_{p}$, we cannot conclusively state that the EEP is violated.
This example shows the variety of properties of astronomical objects. For this simple time lag between different particles, different underlying reasons exist. One should seek more independent methods to obtain different results on the testing of the EEP to overcome the bias (errors) of a special source or of a special method.

Finally, if the non-zero value of $\Delta \gamma_{p}$ is real, (not because of the intrinsic fact of the source nor other trite origins), there are still several complex possibilities. Carroll (1998) suggested that quintessence could rotate the polarization state of radiation from distant sources. Lorentz invariance violation induced vacuum birefringence can introduce a polarization rotation (Fan et al. 2007; Shao \& Ma
2011). Pogosian \& Wyman (2008) considered the contribution of the cosmic strings on the B-mode of cosmic microwave background, as cosmic strings are also able to rotate the polarization angle. Conversely, the B-mode may also place a constraint on the EEP. To distinguish among all of these complex candidates of polarization rotation mechanisms, one should investigate further. With more sources of polarization angles obtained, one may consider the dependence of $\Delta \gamma_{p}$ on the directions, and distances of different sources, which will help to demonstrate the origin of $\Delta \gamma_{p} \neq 0$.

\section*{Acknowledgments}
We thank the editor and the anonymous referee for their valuable comments.
We also thank Xue-Feng Wu, Luis Lehner, Wei-Hua Lei, Fei-Fei Wang, Jun-Jie Wei and Sascha Trippe for helpful discussion.
This research was supported in part by Perimeter Institute for Theoretical Physics. Research at Perimeter Institute is supported by the Government of Canada through the Department of Innovation, Science and Economic Development Canada and by the Province of Ontario through the Ministry of Economic Development, Job Creation and Trade.
This work was supported by the National Natural Science Foundation
of China (Grant Nos. 11703015, U1738132), the China Postdoctoral Science Foundation (Grant
No. 2017M612233), the Natural Science Foundation of Shandong Province (Grant No. ZR2017BA006).

%*******************************************************************************************

%*******************************************************************************************************************

\clearpage
\begin{figure}
 \includegraphics[angle=0,scale=0.5]{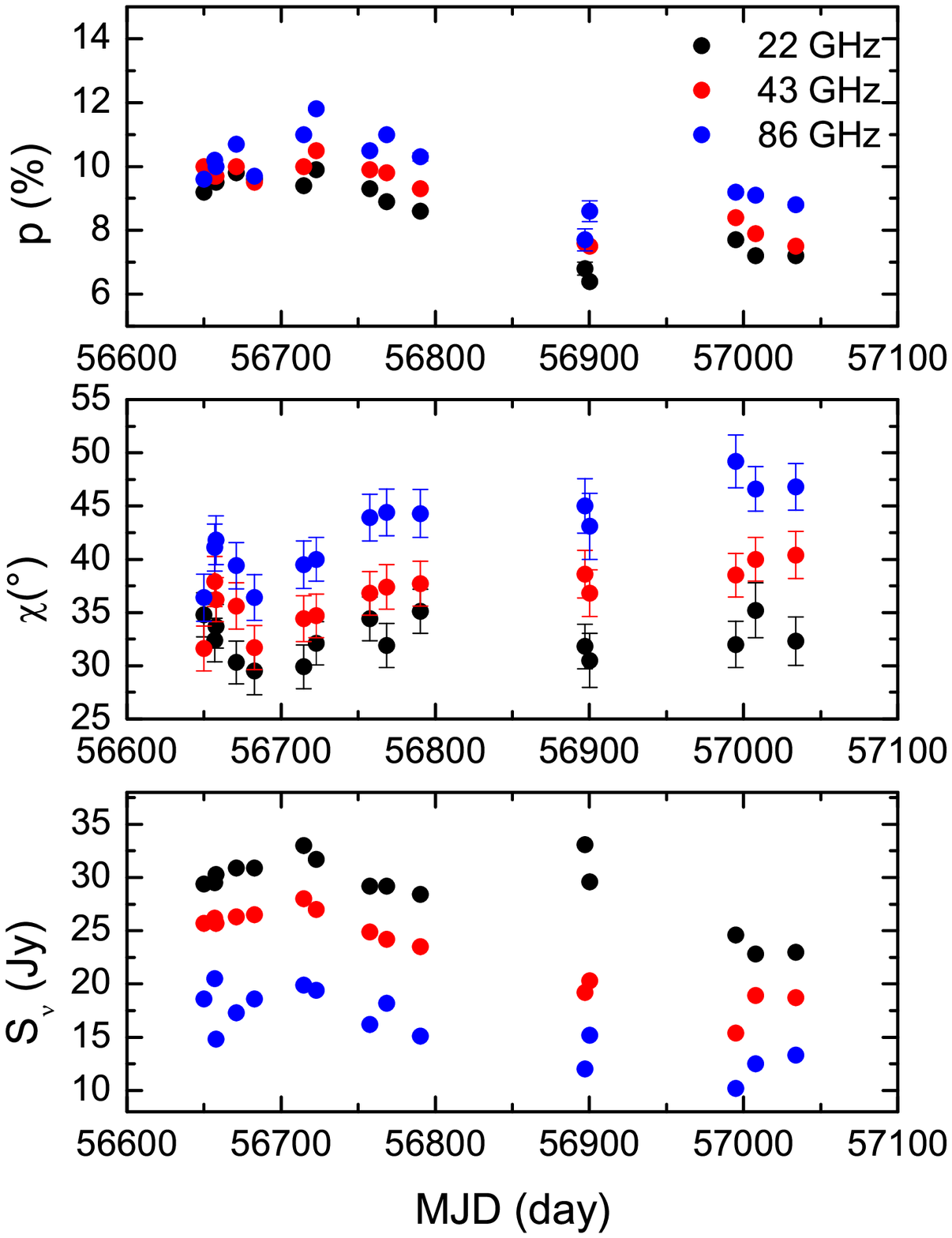}
 \caption{15-point light curves of 3C 279 for degree of linear polarization (top), linear
polarization angle (middle), and flux density (bottom), at 22 GHz (black), 43 GHz (red), and
86 GHz (blue) from December 2013 to March 2015. Data are from Table 2 of Kang et al. (2015).}
\label{fig:data}
\end{figure}

%*******************************************************************************************************************

\clearpage
\begin{center}
\begin{figure} \label{fig:fit2}
 \includegraphics[angle=0,scale=0.5]{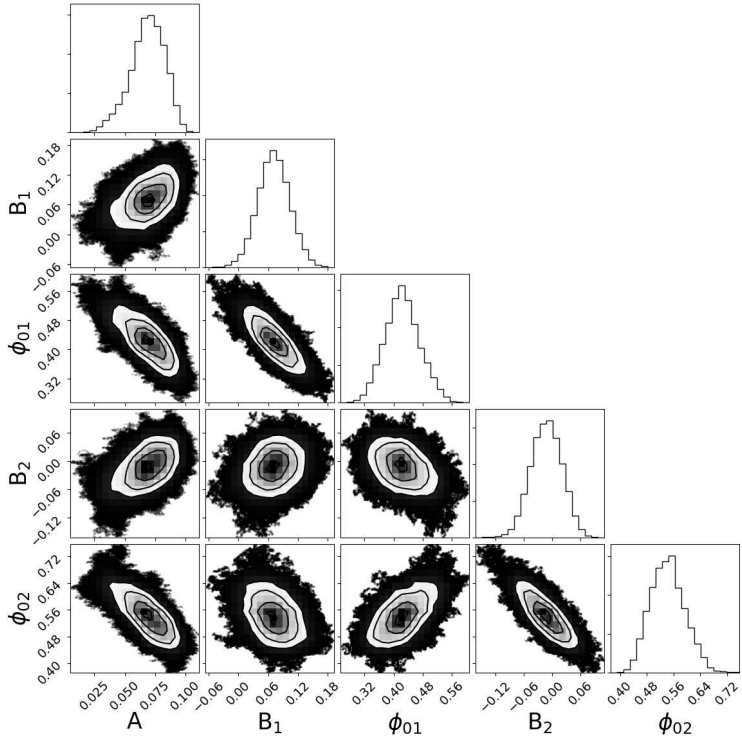}
 \caption{Contour figure of MHMC fitting results for A, $B_1$, $\phi_{01}$, $B_2$, and $\phi_{02}$. The other parameters are added in the Appendix with the contour plots. Values of all 31 fit parameters are shown in Table \ref{tab:2}.}
\end{figure}
\end{center}

%*******************************************************************************************************************
\clearpage
\begin{figure} \label{fig:fit1}
 \includegraphics[angle=0,scale=0.4]{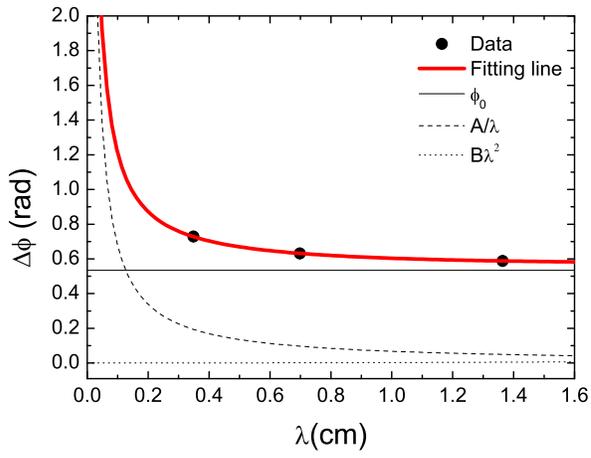}
 \caption{Best-fit result of the third group of data (black dots) with different observed linear polarization angles and multi-wavelength observation for blazar 3C 279. Red line is fit line using Eq. (6), and black lines (solid, dashed, and dotted) are the three different components of Eq. (6), respectively. }
\end{figure}

%*******************************************************************************************************************
\clearpage
\begin{figure} \label{fig:fit2}
 \includegraphics[angle=0,scale=0.35]{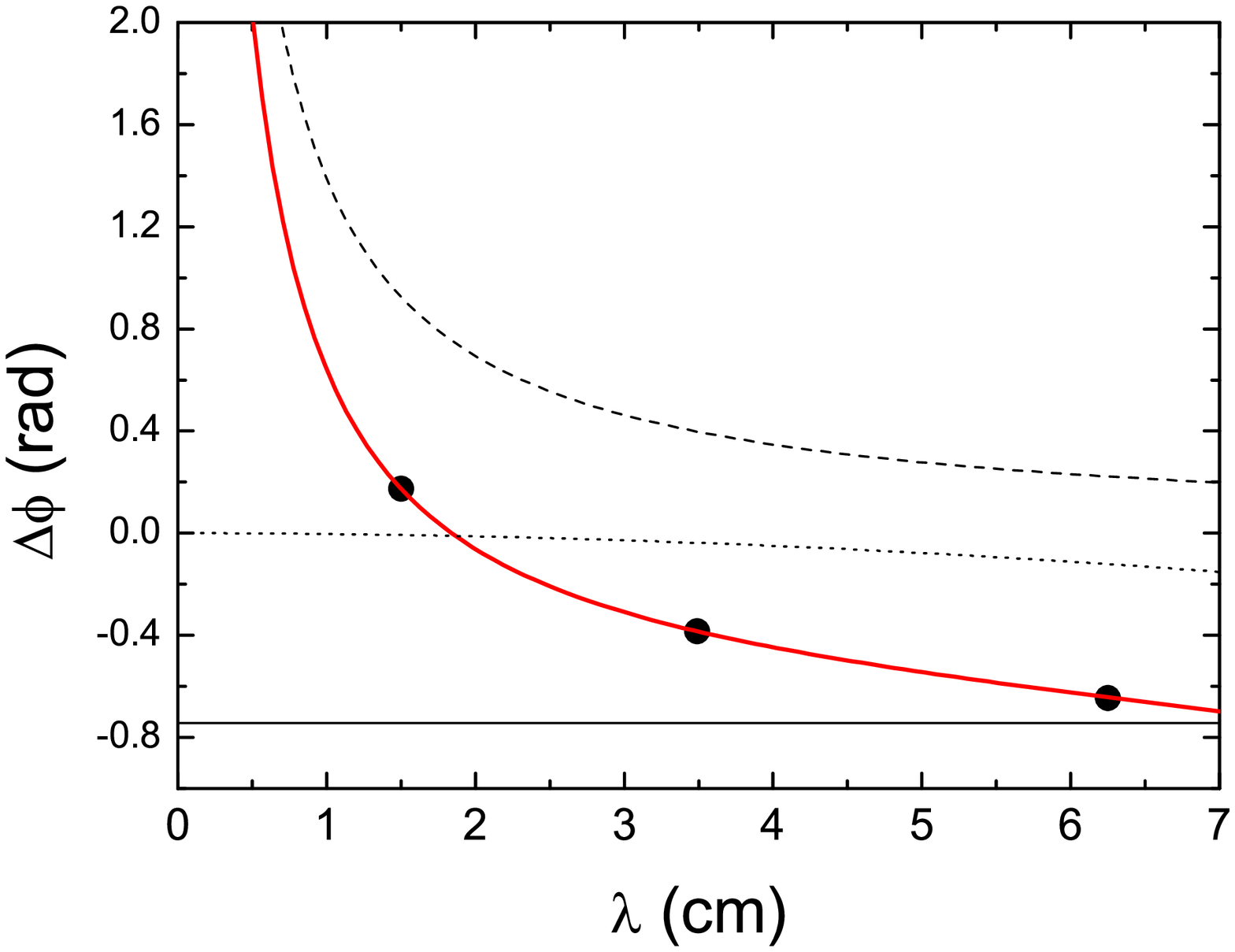}
 \includegraphics[angle=0,scale=0.35]{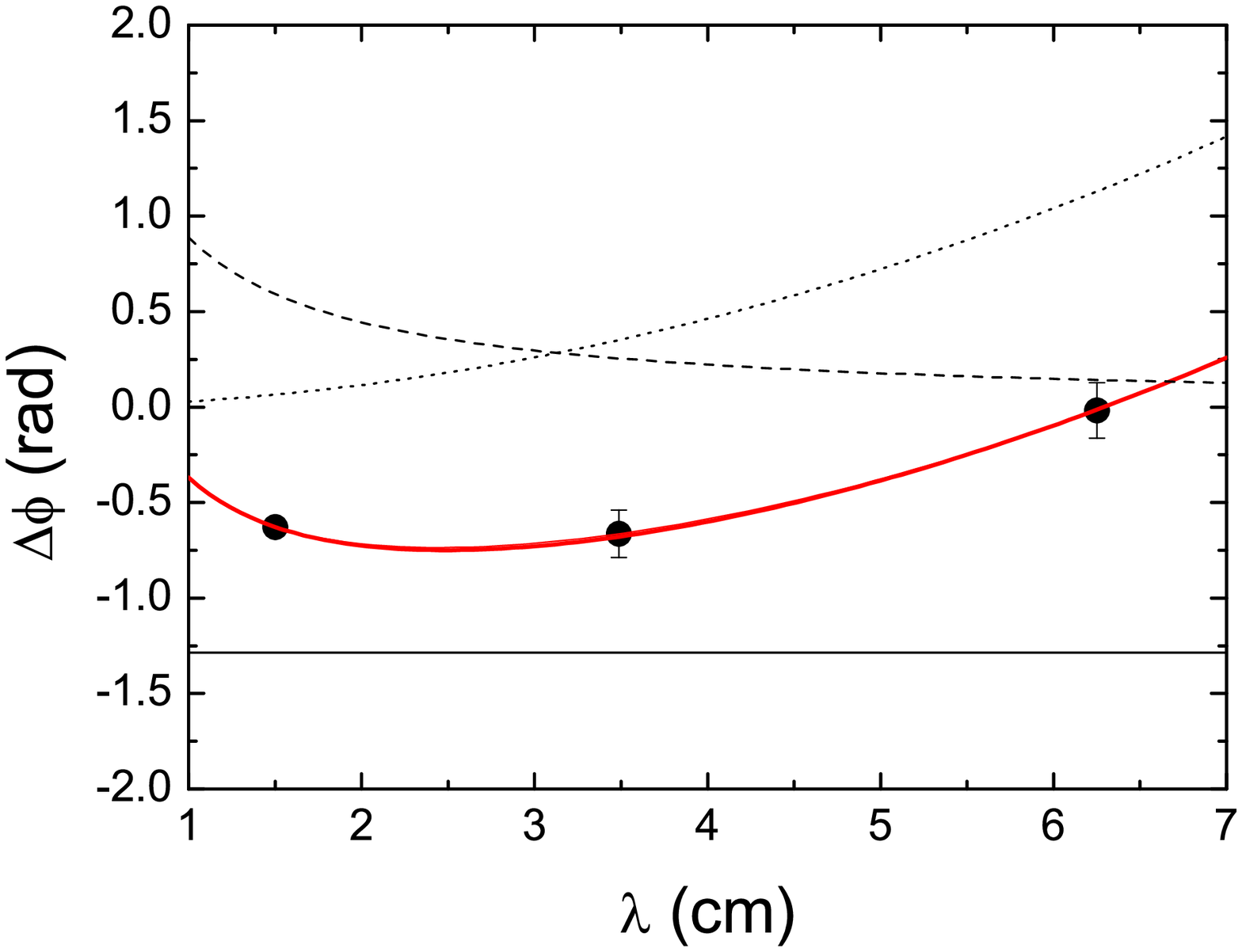}
 \caption{Similar to Figure 3, but for blazars PKS 0008-307 and PKS 0008-264. Fit parameters are presented in Table 2. }
\end{figure}

%*******************************************************************************************************************
\clearpage
\begin{center}
\begin{table}
\caption{Results of multi-frequency polarization observations for 3C 279.
Data are from Table 2 of Kang et al. (2015). Plotted in the second panel of Figure 1.}
\label{tab:1}
\begin{tabular}{lcccccc}
\hline
 & 22~GHz & 43~GHz & 86~GHz &\\
\hline
MJD & $\phi_{\rm obs}(^\circ)$ &$\phi_{\rm obs}(^\circ)$ &$\phi_{\rm obs}(^\circ)$ &\\
  \hline
56649.86	&	34.8	$\pm$	2.07	&	31.6	$\pm$	2.11	&	36.4	$\pm$	2.21	&\\
56656.81	&	32.4	$\pm$	2.05	&	37.9	$\pm$	2.35	&	41.1	$\pm$	2.21	&\\
56657.83	&	33.7	$\pm$	2.04	&	36.2	$\pm$	2.08	&	41.8	$\pm$	2.29	&\\
56670.77	&	30.3	$\pm$	2.01	&	35.6	$\pm$	2.17	&	39.4	$\pm$	2.18	&\\
56682.78	&	29.5	$\pm$	2.24	&	31.7	$\pm$	2.07	&	36.4	$\pm$	2.15	&\\
56714.65	&	29.9	$\pm$	2.04	&	34.4	$\pm$	2.15	&	39.5	$\pm$	2.22	&\\
56722.66	&	32.1	$\pm$	2.02	&	34.7	$\pm$	2.05	&	40.0	$\pm$	2.07	&\\
56757.55	&	34.4	$\pm$	2.06	&	36.8	$\pm$	2.06	&	43.9	$\pm$	2.19	&\\
56768.52	&	31.9	$\pm$	2.08	&	37.4	$\pm$	2.09	&	44.4	$\pm$	2.20	&\\
56790.44	&	35.1	$\pm$	2.08	&	37.7	$\pm$	2.11	&	44.3	$\pm$	2.27	&\\
56897.20	&	31.8	$\pm$	2.11	&	38.6	$\pm$	2.22	&	45.0	$\pm$	2.56	&\\
56900.17	&	30.5	$\pm$	2.53	&	36.8	$\pm$	2.19	&	43.1	$\pm$	3.10	&\\
56994.90    &   32.0	$\pm$	2.16	&	38.5	$\pm$	2.06	&	49.2	$\pm$	2.49	&\\
57007.88	&	35.2	$\pm$	2.57	&	40.0	$\pm$	2.06	&	46.6	$\pm$	2.09	&\\
57033.78	&	32.3	$\pm$	2.28	&	40.4	$\pm$	2.21	&	46.8	$\pm$	2.18	&\\
 \hline
\end{tabular}
\end{table}
\end{center}

%*******************************************************************************************************************

\begin{center}
\begin{table}
\caption{Fit results of 15 groups data for 3C 279 and two other blazars with multi-wavelength linear polarization observations. }
\label{tab:2}
\begin{tabular}{lcccccc}
\hline
 Sources &A (cm) & B ($cm^{-2}$) & $\phi_{\rm 0} \,(rad)$&\\
  \hline
3C 279&0.068	$\pm$	0.012	&	0.069	$\pm$	0.031	&	0.427	$\pm$	0.044	\\
		&	&	-0.013	$\pm$	0.031	&	0.542	$\pm$	0.045	\\
		&	&	0.003	$\pm$	0.030	&	0.534	$\pm$	0.042	\\
		&	&	-0.015	$\pm$	0.030	&	0.510	$\pm$	0.043	\\
		&	&	0.012	$\pm$	0.033	&	0.444	$\pm$	0.045	\\
		&	&	-0.014	$\pm$	0.030	&	0.501	$\pm$	0.043	\\
		&	&	0.005	$\pm$	0.028	&	0.503	$\pm$	0.041	\\
		&	&	-0.008	$\pm$	0.031	&	0.562	$\pm$	0.045	\\
		&	&	-0.042	$\pm$	0.032	&	0.584	$\pm$	0.046	\\
		&	&	-0.003	$\pm$	0.030	&	0.568	$\pm$	0.043	\\
		&	&	-0.050	$\pm$	0.032	&	0.599	$\pm$	0.047	\\
		&	&	-0.045	$\pm$	0.037	&	0.566	$\pm$	0.048	\\
		&	&	-0.074	$\pm$	0.030	&	0.640	$\pm$	0.042	\\
		&	&	-0.029	$\pm$	0.035	&	0.619	$\pm$	0.045	\\
		&	&	-0.067	$\pm$	0.033	&	0.637	$\pm$	0.044	\\
 \hline

PKS 0008-307 & 1.388$\pm$	0.064		&	-0.0031	$\pm$	0.0014	&	-0.744$\pm$	0.033		\\
PKS 0008-264 & 0.888$\pm$	0.519  	    &	0.0289	$\pm$	0.0091    &	-1.285$\pm$	0.365	    \\		
 \hline
\end{tabular}
\end{table}
\end{center}
%*******************************************************************************************************************

%*******************************************************************************************************************

\clearpage
\appendix

\section{Contour plots for the other fit parameters}
For clearer and more complete about the presentation of the fit results of 3C 279, we presented the other parameters of the contour plots in Figure A1.
\begin{center}
\begin{figure} \label{fig:fit2}
 \includegraphics[angle=0,scale=0.35]{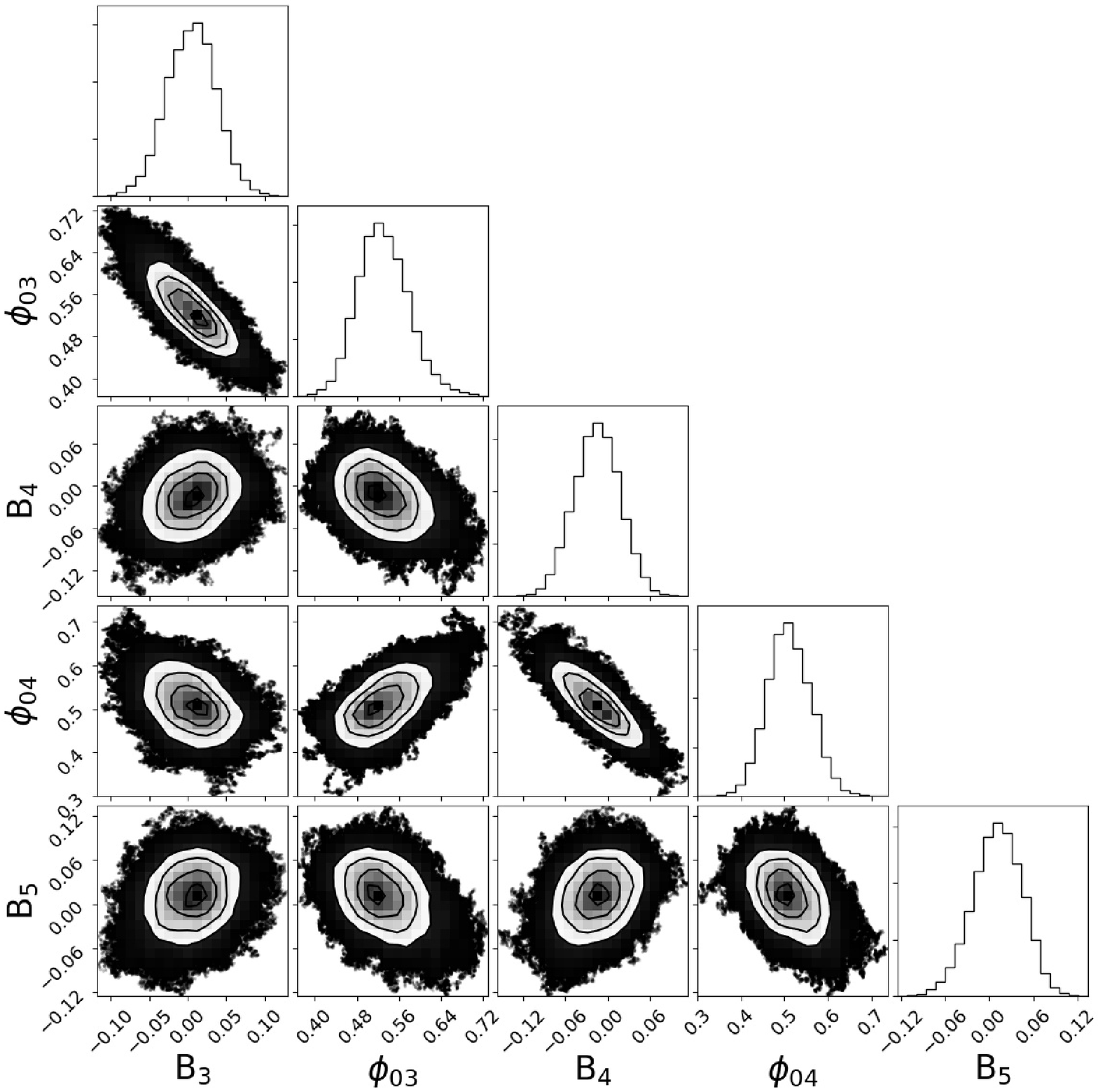}
  \includegraphics[angle=0,scale=0.35]{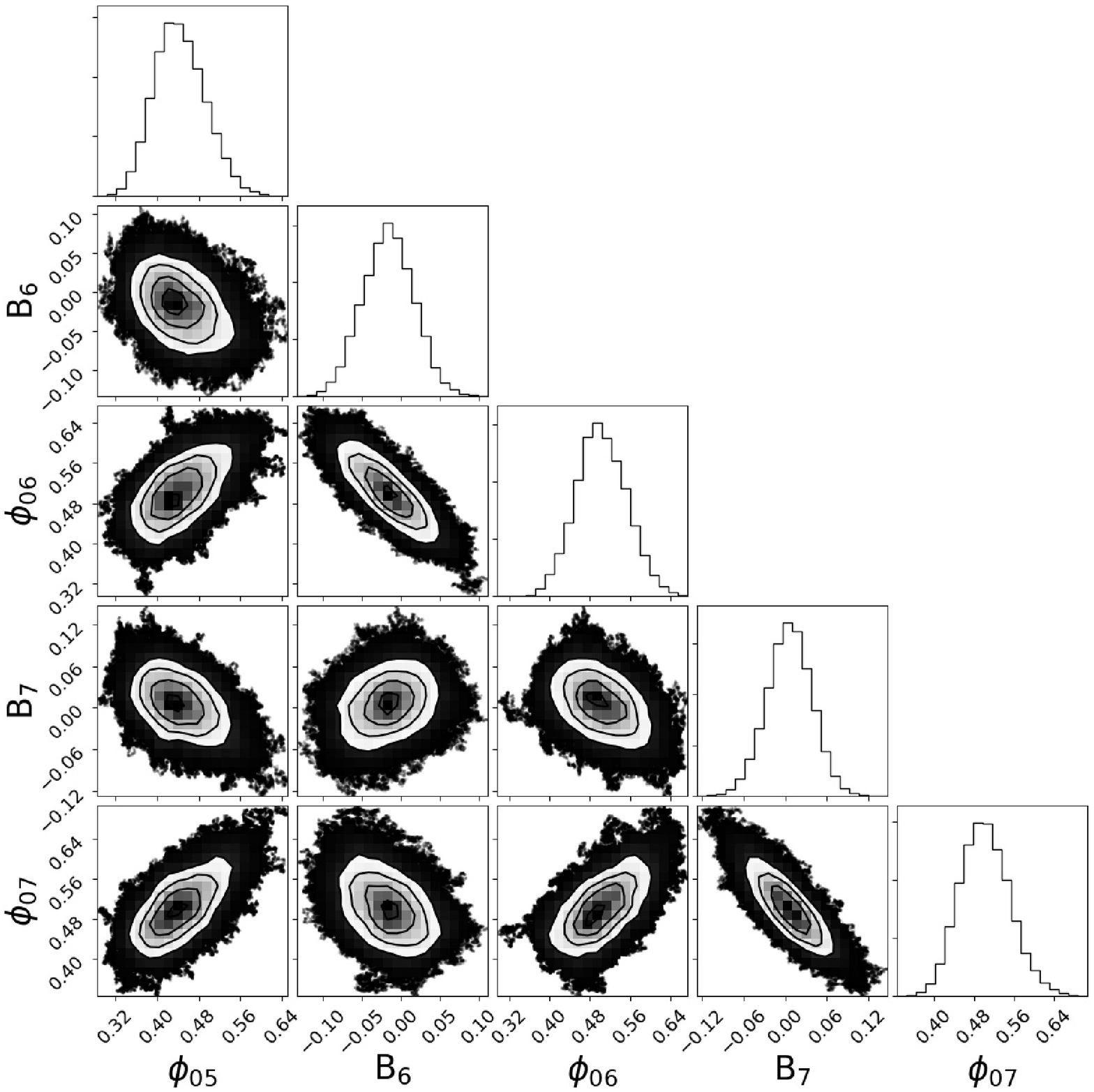}
   \includegraphics[angle=0,scale=0.35]{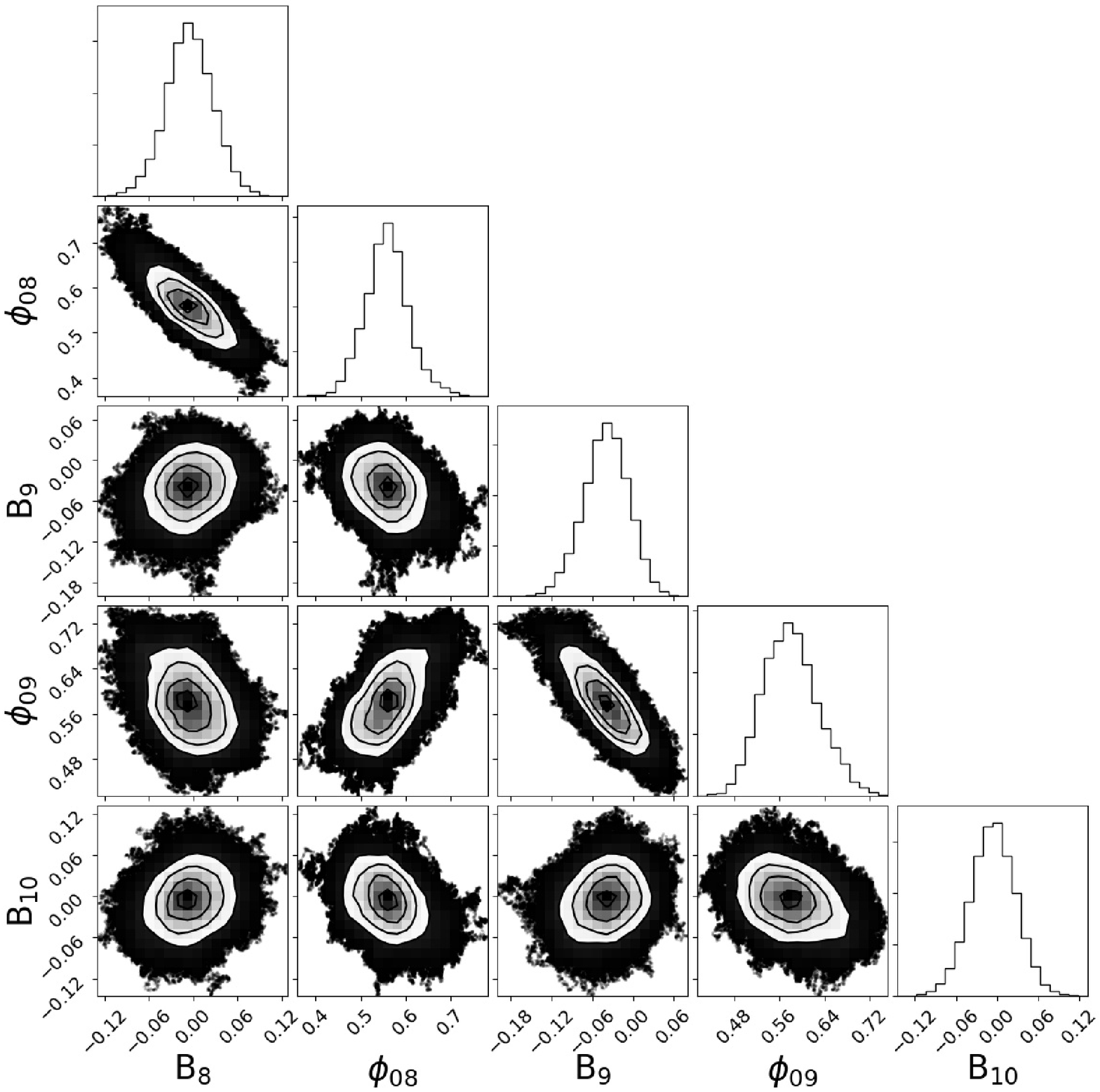}
    \includegraphics[angle=0,scale=0.35]{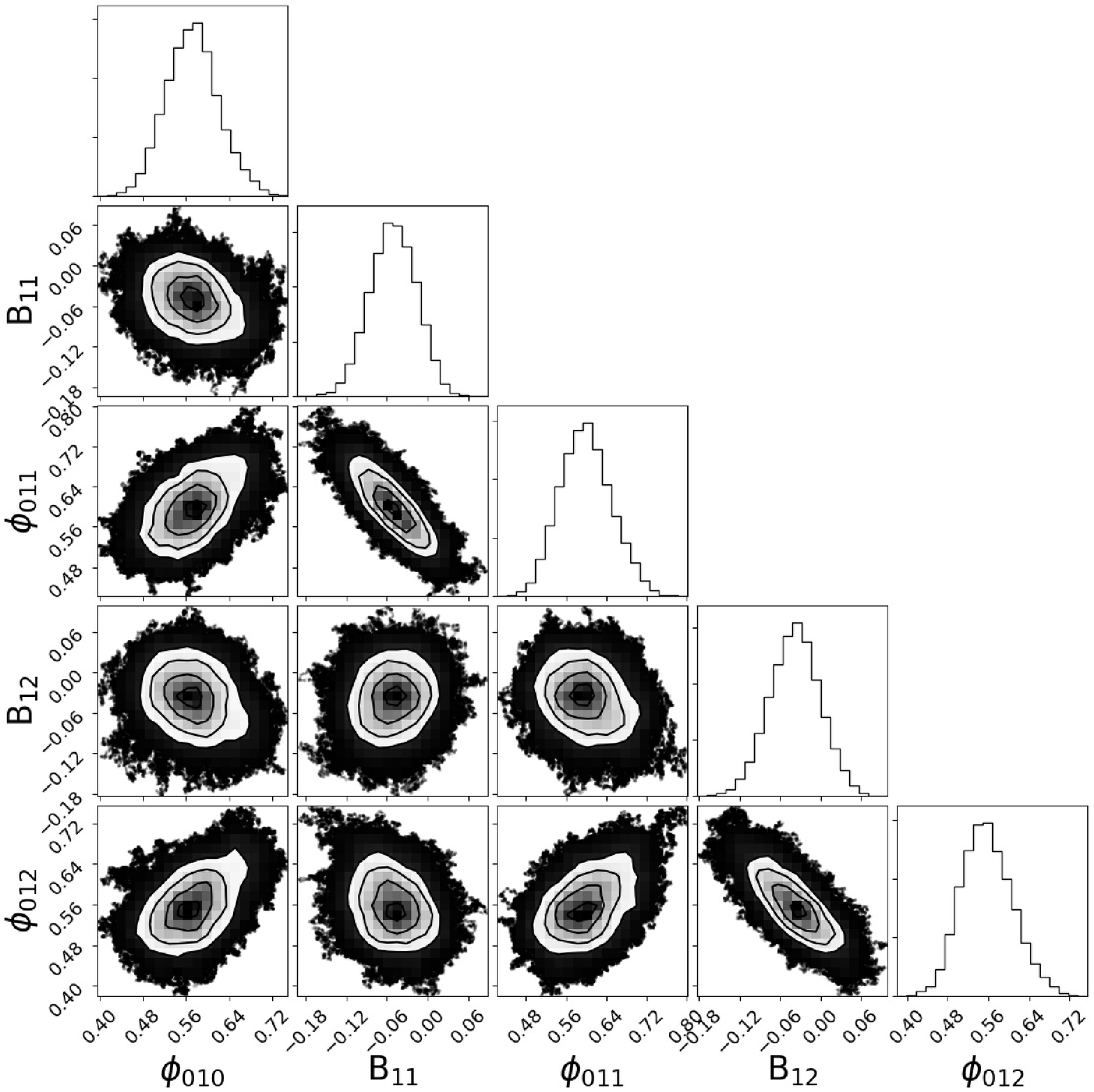}
     \includegraphics[angle=0,scale=0.35]{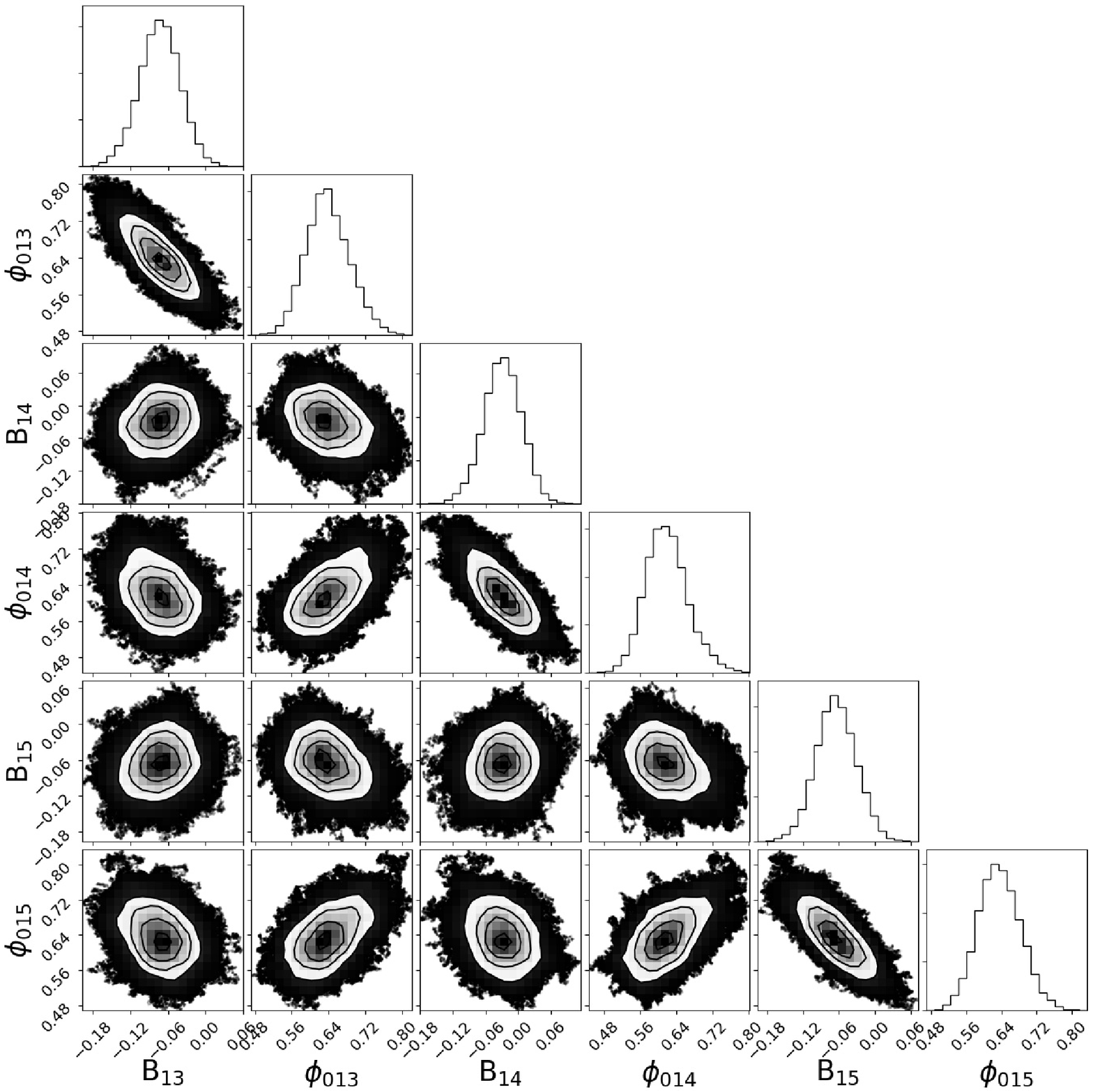}
 \caption{The other parameters of contour plots for 3C 279. }
\end{figure}
\end{center}

\end{document}